\documentclass[a4paper,10pt]{article}
\usepackage{amsmath,amssymb,epsfig,graphicx} 
\usepackage{setspace}

\begin{document}


\title{Simulation of $Z$-boson $p_{T}$ spectrum at LHC and Tevatron 
using GR@PPA}

\author{Shigeru Odaka\\
High Energy Accelerator Research Organization (KEK)\\
1-1 Oho, Tsukuba, Ibaraki 305-0801, Japan\\
E-mail: \texttt{shigeru.odaka@kek.jp}}

\date{}

\maketitle

\begin{abstract}
Predictions from the GR@PPA event generator 
concerning the transverse-momentum ($p_{T}$) spectrum of $Z$ bosons 
are compared with recent measurements at LHC and Tevatron.
The simulation results are in reasonable agreement with the measurements, 
although marginal discrepancies are observed in high-$p_{T}$ regions.
The principal agreements imply that the leading-order simulation 
with a primitive parton shower based on the leading-logarithmic 
approximation still provides a reasonable description of the transverse 
motion of the hard-interaction system in hadron collisions, 
without the need to introduce noble techniques to incorporate 
higher-order corrections.
\end{abstract}

\section{Introduction}
\label{sec:intro}

Hard interactions in hadron collisions, such as heavy particle productions, 
are associated with initial-state hadronic activities 
which induce a transverse motion of the hard-interaction system. 
Since the transverse motion affects the detection capability of experiments, 
a precise understanding of the activities is necessary 
not only for accurate measurements of known processes 
but also for obtaining insights into new physics, 
particularly for discovering the production of new heavy particles 
which may decay to final states including invisible particles. 
The measurement of $Z$-boson production provides a unique opportunity 
for studying these initial-state hadronic activities 
because the production kinematics can be unambiguously measured 
if the production is tagged by a pair of electrons or muons from its decay.

The dominant part of the hadronic activities that may affect the transverse 
motion of the hard interaction system is considered to be caused by 
higher-order strong-interaction (QCD) effects 
which can be evaluated perturbatively.
The higher-order effects are dominated by those terms divergent at the limit 
where partons (light quarks and gluons) are radiated collinear to the incident 
beam directions.
These divergent terms can be factorized and can be added to all orders of the 
QCD coupling, $\alpha_{s}$, to produce finite corrections.
Thus, they are expected to be independent of the details of hard interactions.
The $Q^{2}$ evolution in parton distribution functions (PDF) is 
an implementation of this all-order summation for the longitudinal components. 

The parton shower (PS) is a Monte Carlo (MC) implementation 
of the all-order summation. 
PS simulates transverse activities, which are ignored in PDFs, 
as well as the longitudinal evolution.
Hence, MC event generators implementing PS are indispensable tools 
for evaluating the detection efficiency and acceptance of experiments 
at hadron collisions.
The theoretical basis is identical for PS and the evolution in PDF.
Although the theory is not ambiguous, 
the actual implementation of PS includes substantial ambiguities 
because the introduction of a model is necessary for determining 
the parton-branch kinematics.
A certain model is also necessary for simulating non-perturbative effects.
Therefore, there may be a substantial variation in the predictions from 
different event generators.
Furthermore, for technical reasons, the approximation to the higher-order 
QCD effects is limited to the leading-logarithmic (LL) order in PS 
available at present.
Higher-energy or higher-precision measurements may reveal the applicable 
limit of the approximation.
Hence, we need to repeat studies to examine the reliability of 
event generators and PS 
every time when new data become available from experiments.

GR@PPA 2.8~\cite{Odaka:2011hc,Odaka:2012da} 
is a program package including leading-order (LO) event generators for
single and double weak-boson production in hadron collisions.
Although all matrix elements (ME) are evaluated at LO, 
the event generators can reproduce weak-boson production kinematics 
in the entire phase space by combining non-radiative and radiative 
processes with a jet-matching method.
The package includes PS programs in order to ensure sufficient performance 
of the matching method.
The included PS simulates the higher-order QCD effects down to 
$Q^{2} = (5 {\rm ~GeV})^{2}$.
Further small-$Q^{2}$ phenomena, including non-perturbative effects, can be 
simulated by feeding the generated events to PYTHIA~\cite{Sjostrand:2006za}.
The PS in GR@PPA preserves primitive features of the LL approximation.
It does not implement any noble techniques to import corrections 
from higher-order approximations, such as the angular ordering.
The kinematics model is also very primitive~\cite{Odaka:2007gu,Odaka:2009qf}.
It strictly follows fundamental predictions from the massless approximation.
Therefore, the comparison using this PS should provide a good test bench 
for probing the capability of the LL approximation.

We have shown in a previous study~\cite{Odaka:2009qf} 
that the GR@PPA event generator 
reproduces the transverse momentum ($p_{T}$) spectrum of $Z$ bosons 
produced at FNAL Tevatron, proton-antiproton ($p\bar{p}$) collisions 
at the center-of-mass energy ($\sqrt{s}$) of 1.8 and 1.96 
TeV, with good precision.
Recently, the ATLAS experiment has published 
their measurement result~\cite{Aad:2011gj} at CERN LHC, 
proton-proton ($pp$) collisions at $\sqrt{s}=$ 7 TeV.
In this article, we examine the performance of the GR@PPA event generator 
and its PS by comparing the simulation results with the ATLAS measurement.
Comparisons are also made with the latest $Z \rightarrow e^{+}e^{-}$ 
and $Z \rightarrow \mu^{+}\mu^{-}$ results from the D0 experiment 
at Tevatron.
Although the $Z \rightarrow e^{+}e^{-}$ result from D0 has already been used 
in the previous study, 
the study is extended to more details in this article.
The simulations are carried out by using the GR@PPA 2.8.3 
update release~\cite{Odaka:2012da} through the present study.

The GR@PPA event generator uses a forward-evolution PS (QCDPS) 
for simulating the initial-state QCD activities as the default.
All PS programs in GR@PPA are based on the LO splitting 
functions with $\alpha_{s}$ at the 1-loop approximation.
Therefore, in the default setting, we can use only those PDFs which are 
based on the same approximation.
Otherwise, the QCD evolution of QCDPS leads us to parton distributions 
which are different from those in the selected PDF at large $Q^{2}$ 
relevant to the interested hard interactions.
However, it is known that the LO QCD evolution necessarily gives a 
substantial deficit of quark densities in medium $x$ ($\sim 0.01 - 0.1$) 
regions relevant to weak-boson productions 
at high-energy hadron colliders~\cite{Sherstnev:2007nd}.
It is suggested that it would be better to use 
next-to-leading order (NLO) PDFs even with LO event generators 
to obtain reasonable predictions.
However, NLO PDFs give a small gluon density at small $x$ ($\lesssim 10^{-3}$).
This is not suitable for LO event generations for charm and bottom quark 
productions, and the underlying-event simulation~\cite{Sherstnev:2007nd}.

The so-called modified leading-order (LO*) PDFs have been proposed 
to overcome these difficulties~\cite{Sherstnev:2007nd,Lai:2009ne}. 
Maintaining a high density of gluons at small $x$, 
they significantly enhance the quark densities in medium-$x$ regions 
by relaxing the momentum sum rule at small $Q^{2}$ 
where the evolution originates.
In addition, it is also known that the 1-loop $\alpha_{s}$ approximation 
is not sufficient for reproducing the evolution at small-$Q^{2}$ regions.
The application of the 2-loop $\alpha_{s}$ gives us 
apparently better fits to small-$Q^{2}$ DIS data from HERA experiments 
even with LO PDFs.
For this reason, many of the recent PDFs are using the 2-loop $\alpha_{s}$ 
even if the evolution is at the leading order.

In the present study, we use LO* and NLO PDFs with the GR@PPA event generator.
Although the $Q^{2}$ evolution is performed based on 
the LO approximation in LO* PDFs, 
the evolution in NLO PDFs is explicitly different from that in GR@PPA.
A backward-evolution PS (QCDPSb) is provided in GR@PPA for adopting 
such PDFs based on different assumptions.
In order to make the basic setup common to all simulations, 
we use QCDPSb for the initial-state PS through the study.
Most of the applied PDFs employ the 2-loop $\alpha_{s}$ for the evolution.
Therefore, there may be a mismatch also in the $\alpha_{s}$ approximation.
We have improved the definition of $\Lambda_{\rm QCD}$ in QCDPSb 
in order to perform a consistent evolution in such cases~\cite{Odaka:2012da}.

The remainder of this article is organized as follows. 
The simulation results are compared with the measurement by ATLAS 
in Sec.~\ref{sec:atlas}. 
The simulations with NLO PDFs are discussed along with 
those using LO* and LO PDFs.
Comparison with the D0 data is carried out in Sec.~\ref{sec:d0}, 
and the discussions are concluded in Sec.~\ref{sec:concl}.

\section{Comparison with the ATLAS measurement}
\label{sec:atlas}

\begin{figure}[tp]
\begin{center}
\includegraphics[scale=0.65]{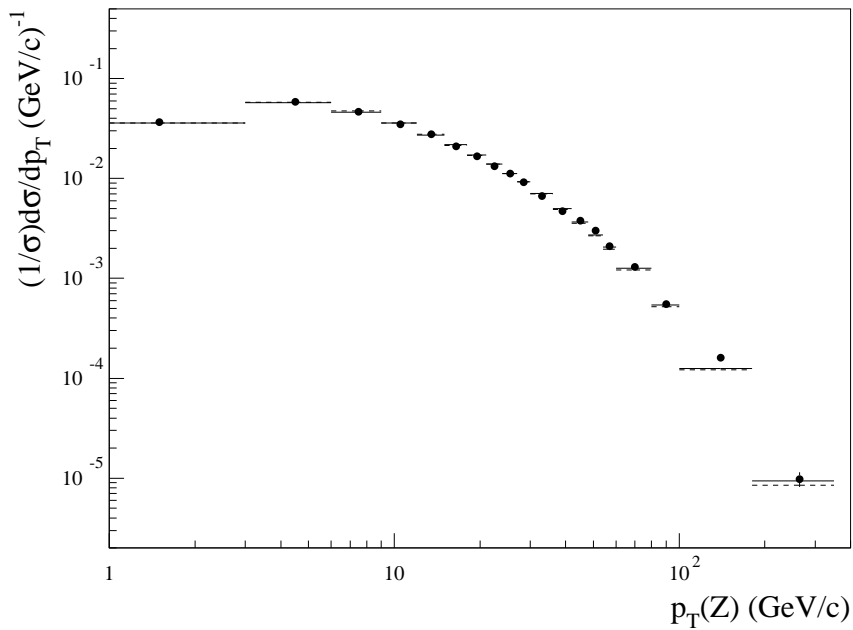}
\caption{\label{fig:atlas_xsec}
Normalized $p_{T}$ distribution of $Z$ bosons in the 7-TeV LHC condition. 
The measurement result from ATLAS (filled circles) is compared with 
the GR@PPA simulations using MRST2007lomod (solid bars) 
and CT09MC1 (dashed bars) for PDF.
}
\end{center}
\end{figure}

Since the main use of MC event generators is for the evaluation of 
the acceptance and efficiency of measurements, 
the relative shape of kinematical distributions is more important 
than the absolute value of the cross sections.
Hence, we examine the $p_{T}$ distribution normalized to the total 
cross section, $(1/\sigma)d\sigma/dp_{T}$, in the present study.
Figure~\ref{fig:atlas_xsec} compares the GR@PPA simulation results 
with the ATLAS measurement~\cite{Aad:2011gj} in this quantity.
Two simulation results are presented: 
one with the MRST2007lomod PDF~\cite{Sherstnev:2007nd} 
and the other with CT09MC1~\cite{Lai:2009ne}.
The simulations were carried out for the 7-TeV LHC condition, 
$pp$ collisions at $\sqrt{s}=$ 7 TeV,
by setting the renormalization scale ($\mu_{R}$) and 
factorization scale ($\mu_{F}$) equal to the $Z$-boson mass, 
$m_{Z} = 91.19 {\rm ~GeV}/c^{2}$.
The energy scale for the final-state PS was also set to the same value.
The PDFs were referred to by using the LHAPDF library~\cite{Whalley:2005nh}.

The events were generated with a constraint on the $Z$-boson invariant mass 
of $66 \leq m_{Z} \leq 116 {\rm ~GeV}/c^{2}$.
The parton showers were applied down to $Q^{2} = (5 {\rm ~GeV})^{2}$ in GR@PPA.
The backward-evolution PS (QCDPSb) was used for the initial state.
As usual, the generated events were passed to PYTHIA 
for simulating lower-$Q^{2}$ phenomena down to the hadron level.
The PYTHIA version 6.425 was used for this simulation.
The default setting in PYTHIA was unchanged except for the setting of 
{\tt PARP(67) = 1.0} and {\tt PARP(71) = 1.0}, 
as in the previous study~\cite{Odaka:2009qf}.
After applying the PYTHIA simulation, 
we constrained the pseudorapidity ($\eta$) and 
transverse momentum ($p_{T}$) of the decay leptons as 
$|\eta| \leq 2.4$ and $p_{T} \geq 20 {\rm ~GeV}$, 
following the signal definition of ATLAS.
These conditions were imposed on the quantities 
before adding the photon-radiation effects in the decay, 
according to the ATLAS definition.

\begin{figure}[tp]
\begin{center}
\includegraphics[scale=0.65]{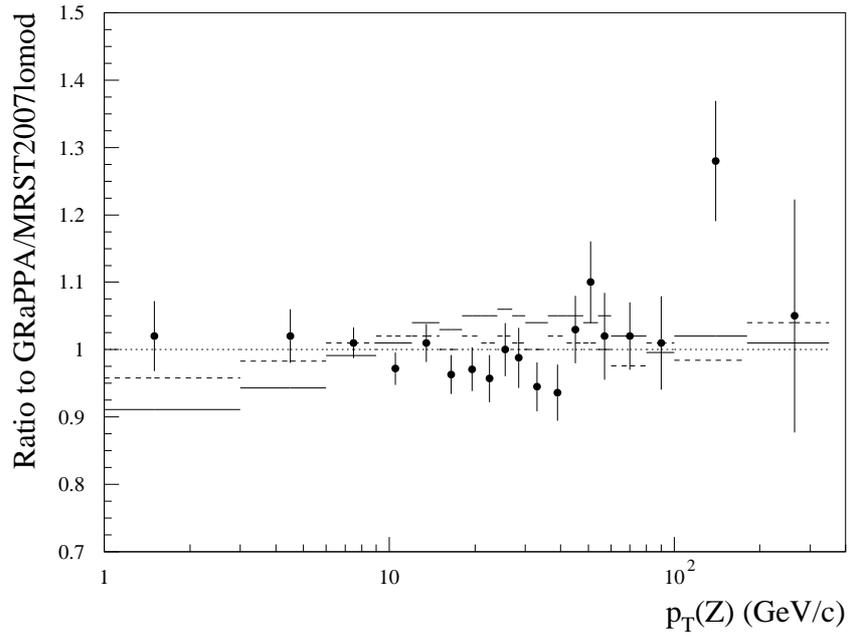}
\caption{\label{fig:atlas_mrst}
The ratio of the ATLAS data to the GR@PPA simulation with the MRST2007lomod PDF.
The solid bars show a test simulation result in which the $\Lambda_{\rm QCD}$ 
value is increased to 0.165 GeV, 
and the dashed bars illustrate the result from the simulation with CTEQ6L1.
}
\end{center}
\end{figure}

\begin{figure}[tp]
\begin{center}
\includegraphics[scale=0.65]{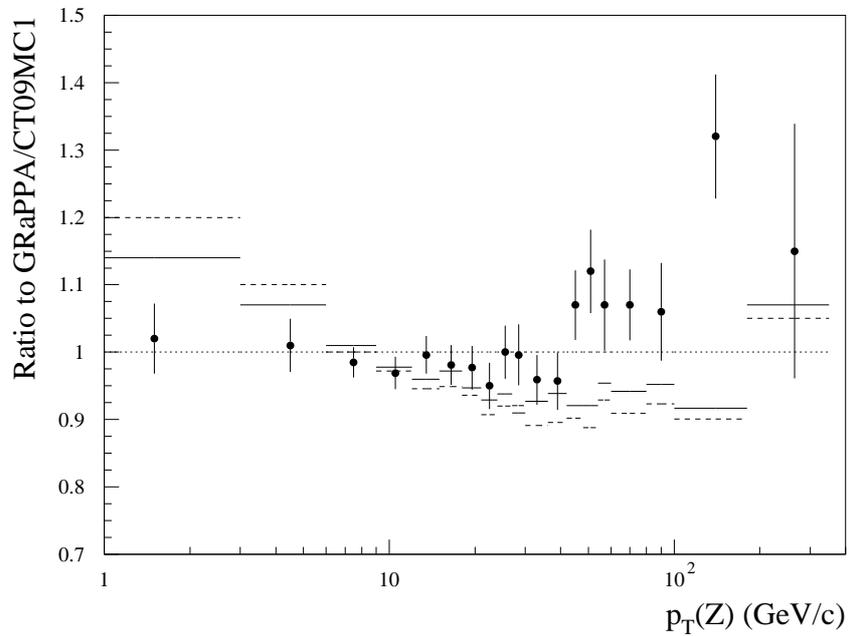}
\caption{\label{fig:atlas_ct09}
The ratio of the ATLAS data to the GR@PPA simulation with the CT09MC1 PDF.
Solid and dashed bars show the simulation results with NLO PDFs, 
MSTW2008nlo and CT10, respectively.
}
\end{center}
\end{figure}

We can see an overall agreement between the simulations and the measurement 
in Fig.~\ref{fig:atlas_xsec}.
However, details are invisible because the results extend to a very wide 
range of approximately four orders of magnitude.
In order to closely examine the details, 
we have plotted the ratio of the data to a simulation result in 
Fig.~\ref{fig:atlas_mrst}.
Here, first of all, we used the simulation with MRST2007lomod 
as the reference, 
because the ATLAS group has stated that the PYTHIA simulation with 
this PDF shows good agreement with their measurement.
We can observe good agreement between the simulation and the data 
within the quoted measurement uncertainty even with the GR@PPA simulation, 
throughout the measurement range except for the second-highest $p_{T}$ bin.

MRST2007lomod is a so-called LO* PDF 
and uses the 2-loop $\alpha_{s}$ for the evolution 
with $\Lambda_{\rm QCD} = 0.241 {\rm ~GeV}$.
This $\Lambda_{\rm QCD}$ value corresponds to the $\alpha_{s}(m_{Z}^{2})$ 
value of 0.119.
Therefore, this is a good test case for our definition 
of $\Lambda_{\rm QCD}$.
GR@PPA adjusts the $\Lambda_{\rm QCD}$ for PS 
so that the used 1-loop $\alpha_{s}$ should reproduce 
the 2-loop $\alpha_{s}$ in PDF as precisely as possible~\cite{Odaka:2012da}.
The adjustment gives a $\Lambda_{\rm QCD}$ value of 0.116 GeV, 
which leads to $\alpha_{s}(m_{Z}^{2}) = 0.123$.
The result in Fig.~\ref{fig:atlas_mrst} implies that this definition 
is reasonable.

As a test, we carried out a simulation with 
$\Lambda_{\rm QCD} = 0.165 {\rm ~GeV}$, 
the default setting in the original GR@PPA 2.8.
This value is adopted in CTEQ6L1~\cite{Pumplin:2002vw}, 
an LO PDF employing the 1-loop $\alpha_{s}$ leading to 
$\alpha_{s}(m_{Z}^{2}) = 0.130$.
The ratio to the current default result is shown with solid bars 
in Fig.~\ref{fig:atlas_mrst}.
We can see a significant discrepancy from the data with this choice.
From this result, we can also find that the $p_{T}$ range can be subdivided 
to three regions: $p_{T} \lesssim 10 {\rm ~GeV}$ ($R_{1}$), 
$10 \lesssim p_{T} \lesssim 100 {\rm ~GeV}$ ($R_{2}$), and 
$p_{T} \gtrsim 100 {\rm ~GeV}$ ($R_{3}$).
In the previous study, we found that the primordial $k_{T}$ simulation 
in PYTHIA determines the shape within $R_{1}$, whereas it does not alter 
the distribution in $R_{2}$ and $R_{3}$.
The results in Fig.~\ref{fig:atlas_mrst} show that the balance between 
the fractions in $R_{1}$ and $R_{2}$ is a good measure for the validation 
of the PS simulation.
The fraction in $R_{2}$ increases as we choose larger $\Lambda_{\rm QCD}$ 
values leading to larger values of $\alpha_{s}$,
and the fraction in $R_{1}$ decreases to compensate for it.
The fraction in $R_{3}$ is nearly independent of these simulations 
because it is predominantly determined by the $Z$ + 1 jet cross section.

The dashed bars in Fig.~\ref{fig:atlas_mrst} show the result from 
the simulation with the CTEQ6L1 PDF.
Of course, $\Lambda_{\rm QCD} = 0.165 {\rm ~GeV}$ is assumed 
with the 1-loop $\alpha_{s}$ in this simulation.
Despite this assumption, 
although a slight enhancement is seen in the $R_{2}$ region, 
the result does not show a large discrepancy from the MRST2007lomod result.
The input parton distributions at small $Q^{2}$ must be appropriately adjusted 
in CTEQ6L1 to give comparable distributions in the relevant $Q^{2}$ range.
Although the forward-evolution PS can be used with this PDF, 
the presented result has been obtained with the backward-evolution PS.
We have confirmed that the application of the forward-evolution PS does not 
significantly alter the result. 

Figure~\ref{fig:atlas_ct09} shows the result in which the GR@PPA simulation 
with another LO* PDF, CT09MC1, is used as the reference.
Contrary to MRST2007lomod, CT09MC1 uses the 1-loop $\alpha_{s}$ 
for the evolution;
thus, $\alpha_{s}(m_{Z}^{2})$ = 0.130 and the $\alpha_{s}$ value is always 
larger than the MRST2007lomod value by about 10\% in the GR@PPA PS.
Despite such a difference, the result in Fig.~\ref{fig:atlas_ct09} is 
nearly identical to the previous result in Fig.~\ref{fig:atlas_mrst}.
This agreement further justifies our method for determining 
the $\Lambda_{\rm QCD}$ value.
However, the high-$p_{T}$ behavior of this result is a little bit strange.

GR@PPA uses the $\alpha_{s}$ value that is given by the selected PDF 
for the calculation of matrix elements.
The high-$p_{T}$ ($\gtrsim 100 {\rm ~GeV}/c$) cross section is determined by 
$Z$ + 1 jet matrix elements which are proportional to $\alpha_{s}$.
Thus, as a naive speculation, the $R_{3}$ fraction of the simulation 
with CT09MC1 should be larger than that of the simulation with MRST2007lomod 
by about 10\%, since we always set $\mu_{R} = m_{Z}$.  
In other words, the ratio of the data shown in Fig.~\ref{fig:atlas_ct09} should 
be smaller than that in Fig.~\ref{fig:atlas_mrst} by about 10\% 
in the region $R_{3}$.
This amount of systematic difference should be visible 
because the statistical error of the simulation is only 3.6\% 
in the highest $p_{T}$ bin and 1.5\% in the second highest bin 
in these simulations. 
However, no such tendency is observed.
This is mainly because the $q \bar{q} \rightarrow Z$ cross section 
of the simulation with CT09MC1 is significantly ($\sim 9\%$) larger than 
the simulation with MRST2007lomod, 
owing to a larger enhancement of the quark density in CT09MC1.

The gluon density contributes to the $Z$ + 1 jet cross section 
through the $qg \rightarrow Z + q$ interaction.
Therefore, if there is no difference in the gluon density, 
the larger quark density should lead to another enhancement in 
the high-$p_{T}$ cross section of the simulation with CT09MC1, 
in addition to the enhancement due to the larger $\alpha_{s}$ value.
Thus, the high-$p_{T}$ cross section of the simulation with CT09MC1 
should be larger than that of the simulation with MRST2007lomod 
by more than 10\% from this naive estimation.
However, the enhancement is only 5\% near $p_{T} = 100 {\rm ~GeV}/c$.
As a result, contrary to the naive expectation, 
the high-$p_{T}$ fraction of the former becomes about 4\% smaller 
than the fraction of the latter.
This implies that the difference in the gluon density must also be substantial. 

As we have seen in the above, 
the studied LO* and LO PDFs provide nearly identical $Z$-boson $p_{T}$ spectra, 
despite there being substantial differences in the setting of the fundamental 
parameters and assumptions between them. 
The parton distributions in the PDFs must have been appropriately adjusted 
through the fits to experimental measurements.
However, this fact does not imply that the $Z$-boson $p_{T}$ spectrum is 
insensitive to the variation in PDF.

It is suggested for many years that it would be better to use 
NLO PDFs even with LO event generators 
to obtain reasonable predictions, 
particularly concerning rapidity distributions.
We have tested such simulations employing recent NLO PDFs.
The results are plotted in Fig.~\ref{fig:atlas_ct09}.
The solid bars illustrate the result obtained with 
the MSTW2008nlo~\cite{Martin:2009iq} PDF, 
and dashed bars with CT10~\cite{Lai:2010vv}.
The two results show a nearly identical behavior and fail to reproduce 
the measurement data.
The fraction in the $R_{1}$ region is too large and, correspondingly, 
too small in $R_{2}$.
We have observed similar results from the simulations 
with the PYTHIA built-in generator.

Similar results have also been obtained in the report from 
ATLAS~\cite{Aad:2011gj} for NLO predictions in which NLO PDFs are employed.
It should be reminded that the $p_{T}$ spectrum is evaluated on the basis of 
the tree-level MEs for $Z$ + 1 jet processes even in NLO calculations.
Among several NLO predictions, 
the RESBOS prediction provides a reasonable explanation of the measurement 
in the ATLAS report.
This is because the compared RESBOS prediction is corrected by using a K-factor 
derived from a next-to-next-to-leading order (NNLO) calculation.
These facts imply that there must be a fundamental mismatch between tree-level 
MEs and NLO PDFs.

\section{Comparison with the D0 measurements}
\label{sec:d0}

\begin{figure}[tp]
\begin{center}
\includegraphics[scale=0.65]{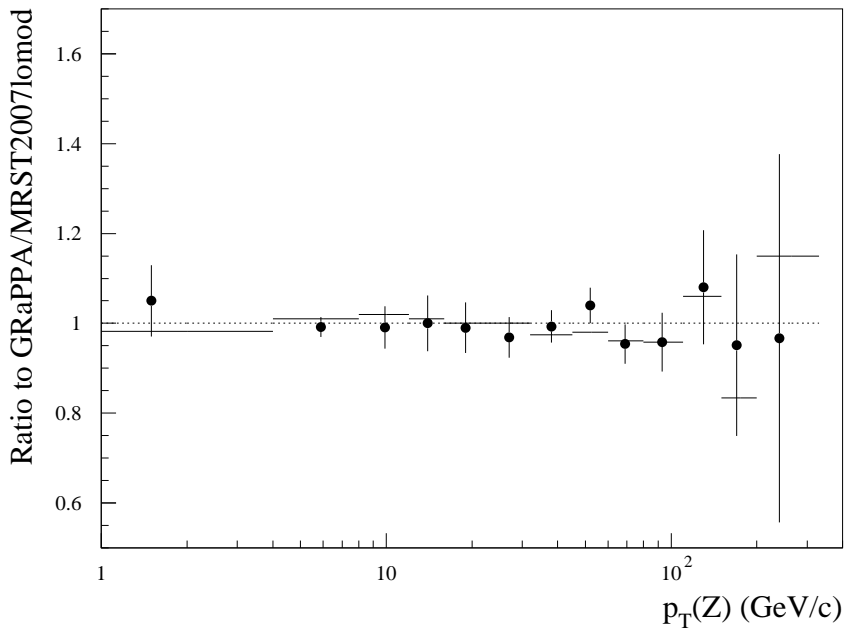}
\caption{\label{fig:d0mumu}
The ratio of the $Z \rightarrow \mu^{+}\mu^{-}$ data from D0 
to the GR@PPA simulation with MRST2007lomod.
The solid bars show the simulation result with the CT09MC1 PDF.
The D0 data are extrapolated to the condition of 
$65 \leq m_{Z} \leq 115 {\rm ~GeV}/c^{2}$ with $|\eta(\mu)| \leq 1.7$ 
using the correction factors presented in their report.
The simulation results are obtained with the same condition.
The muon properties without the photon radiation effect are used to 
define the condition.
}
\end{center}
\end{figure}

In this section, we compare the GR@PPA simulation results with the 
$Z$-boson $p_{T}$ spectrum data from the D0 experiment 
at Tevatron Run-II, 
$p\bar{p}$ collisions at $\sqrt{s} = 1.96 {\rm ~TeV}$.
We first examine the measurement on the $Z \rightarrow \mu^{+}\mu^{-}$ 
channel recently published~\cite{Abazov:2010kn}.
D0 has presented a result for a constraint close to their actual 
detection condition.
The condition that they have defined is not convenient for simulations 
because it is not easy to guarantee the accuracy of the simulation 
for the final-state photon radiation effect.
Instead, we extrapolate the D0 data to the simplified detection condition, 
$65 \leq m_{Z} \leq 115 {\rm ~GeV}/c^{2}$ with $|\eta(\mu)| \leq 1.7$,
using the correction factors provided in their paper.
We have multiplied their measurement data with those factors denoted by 
"$p_{T}^{\mu}$" and "FSR" in Table 1 of the paper.
Thus, we can consider that the constraints are satisfied in the properties 
without the final-state photon radiation effects.
The simulations have been carried out with this assumption.

Figure~\ref{fig:d0mumu} shows the ratio of the D0 data to the GR@PPA 
simulation with MRST2007lomod.
The error bars were evaluated by the quadratic sum of all errors 
presented in the D0 paper.
We can see that the agreement is very good throughout the measurement range.
The solid bars in Fig.~\ref{fig:d0mumu} show the simulation result with 
CT09MC1.
The PDF dependence is very small compared to the measurement error.

\begin{figure}[tp]
\begin{center}
\includegraphics[scale=0.65]{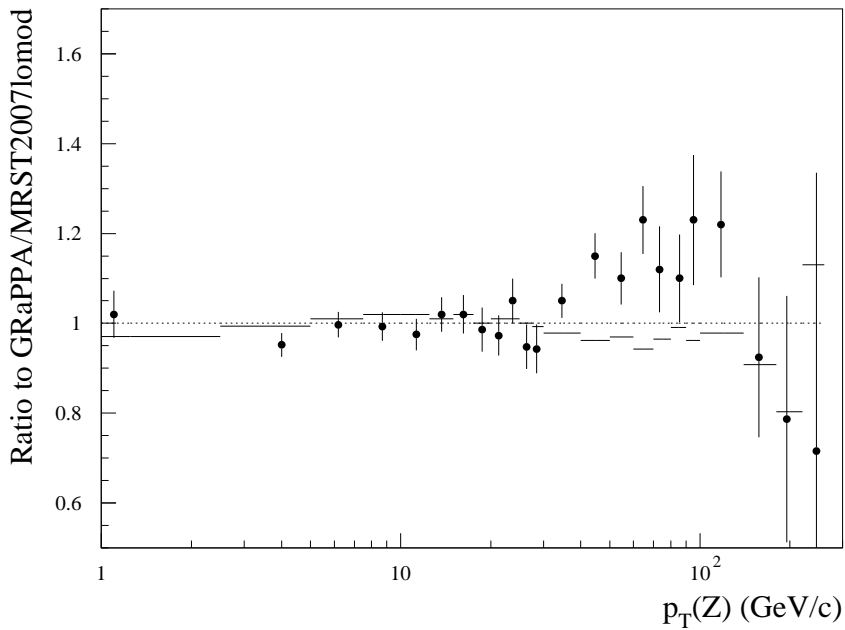}
\caption{\label{fig:d0ee}
The ratio of the $Z \rightarrow e^{+}e^{-}$ data from D0 
to the GR@PPA simulation with MRST2007lomod.
The solid bars show the result with the CT09MC1 PDF.
The simulation results are obtained for the $Z$-boson mass constraint 
of $40 \leq m_{Z} \leq 200 {\rm ~GeV}/c^{2}$ with no cut on the decay 
electrons, according to the definition by D0.
}
\end{center}
\end{figure}

Next, we examine the D0 data on the $Z \rightarrow e^{+}e^{-}$ channel 
published several years ago~\cite{Abazov:2007nt}.
The data presented in the report are extrapolated to those 
with a very loose constraint, $40 \leq m_{Z} \leq 200 {\rm ~GeV}/c^{2}$ 
without any cut on electrons.
Simulations were carried out for this condition.
Figure~\ref{fig:d0ee} shows the ratio of the D0 data to 
the simulation with MRST2007lomod.
The agreement is very good at low $p_{T}$ ($\lesssim 40 {\rm ~GeV}/c$)
whereas a systematic enhancement of the data or deficit of the simulation 
is observed in the $p_{T}$ range of 
$40 \lesssim p_{T} \lesssim 140 {\rm ~GeV}/c$.
The solid bars in Fig.~\ref{fig:d0ee} show the simulation result with CT09MC1.
The PDF dependence is small compared to the measurement error.

The discrepancy in the high $p_{T}$ region described above 
was also observed by D0 in the comparison of their result 
with NLO and resummation predictions~\cite{Abazov:2007nt}. 
A similar enhancement at high $p_{T}$, or a jump of the cross section 
around $p_{T} = 40 {\rm ~GeV}/c$ ($\sim m_{Z}/2$), 
is also seen in the ATLAS result in Figs.~\ref{fig:atlas_mrst} 
and \ref{fig:atlas_ct09}, although the significance is marginal.
On the other hand, no such disagreement is observed 
in the D0 $Z \rightarrow \mu^{+}\mu^{-}$ data shown in Fig.~\ref{fig:d0mumu}.
It is impossible to reproduce these observations simultaneously 
with our simulation.
Although these observations may suggest a defect in the simulation 
or a sign of new physics, 
we also suspect a common problem in the measurements.
The $p_{T}$ spectrum of decay leptons is broadened 
as the $p_{T}$ of $Z$ bosons ($p_{T}(Z)$) increases.
Accordingly, the inefficiency due to the $p_{T}$ cut in the lepton detection 
becomes larger in higher $p_{T}(Z)$ bins.
An underestimation of the efficiency or a shift in the energy scale 
at small $p_{T}$ near the detection threshold may result in 
a systematic enhancement of the cross section at high $p_{T}(Z)$. 
In any case, the observed discrepancies are still marginal compared to 
the measurement errors.
Hence, more precise measurements with improved statistics at LHC are awaited 
in order to clarify the reasons for these observations.

\section{Conclusions}
\label{sec:concl}

$Z$-boson production is a unique place for measuring the initial-state 
hadronic activities in hadron collisions, 
which are considered to be common to all hard interactions.
The PS in GR@PPA, 
which simulates dominant parts of the hadronic activities, 
strictly preserves primitive features of 
the LL approximation of QCD.
Hence, the comparison of its prediction with measurement data provides us 
with a unique opportunity for probing the applicable limit of the 
LL approximation.

In this article, we have compared the predictions from the GR@PPA event 
generator with recent measurements on the $Z$-boson $p_{T}$ spectrum 
at LHC and Tevatron.
We have primarily used recently-proposed LO* PDFs for the simulation, 
in combination with a backward-evolution PS.
The predictions with the MRST2007lomod and CT09MC1 PDFs have been discussed.
Although the approximation order of $\alpha_{s}$ is different between 
the two PDFs, 
an appropriate definition of the strong-interaction coupling $\alpha_{s}$ 
in GR@PPA has made it possible to carry out a consistent simulation.

The simulation results are in good agreement with the measurement data, 
at least in low-$p_{T}$ regions, $p_{T} \lesssim 40 {\rm ~GeV}/c$,
where accurate data are available and 
the multiple radiation by PS predominantly determines the spectrum.
The agreement is well within 5\%.
The difference between the predictions with the two PDFs is smaller 
than this level. 
We have also found that the difference to the prediction 
with the LO PDF is not remarkable in the $p_{T}$ spectrum.
On the other hand, simulations with NLO PDFs 
show significant discrepancies from the measurements.
There must be a fundamental mismatch in the combination of tree-level 
matrix elements with NLO PDFs.

Although the overall shape is consistent, 
the $Z \rightarrow e^{+}e^{-}$ data from D0 at Tevatron show 
a substantial enhancement of more than 10\% with respect to the simulations 
in the $p_{T}$ range of $40 \lesssim p_{T} \lesssim 140{\rm ~GeV}/c$.
A similar enhancement is also seen in the ATLAS data, 
whereas no such enhancement is observed in the $Z \rightarrow \mu^{+}\mu^{-}$ 
data from D0.
The observed discrepancies are still marginal and 
precise measurements with improved statistics at LHC are awaited 
for further discussions.

\section*{Acknowledgments}

This work has been carried out as an activity of the NLO Working Group, 
a collaboration between the Japanese ATLAS group and the numerical analysis 
group (Minami-Tateya group) at KEK.
The author wishes to acknowledge useful discussions with the members, 
especially with Y. Kurihara.



\begin{thebibliography}{10}

\bibitem{Odaka:2011hc}
S.~Odaka and Y.~Kurihara, {\it {GR@PPA 2.8: initial-state jet matching for
weak-boson production processes at hadron collisions}},  
Comput. Phys. Commun. {\bf 183} (2012) 1014; 
arXiv:1107.4467.

\bibitem{Odaka:2012da}
S.~Odaka, {\it {GR@PPA 2.8.3 update}}, arXiv:1201.5702.

\bibitem{Sjostrand:2006za}
T.~Sjostrand, S.~Mrenna, and P.~Z. Skands, {\it {PYTHIA 6.4 Physics and
Manual}},  JHEP {\bf 05} (2006) 026; 
arXiv:hep-ph/0603175.

\bibitem{Odaka:2007gu}
S.~Odaka, Y.~Kurihara, {\it {Initial-state parton shower kinematics for NLO event
generators}}, Eur. Phys. J. C {\bf 51} (2007) 867;
arXiv:hep-ph/0702138.

\bibitem{Odaka:2009qf}
S.~Odaka, {\it {Simulation of $Z$ boson $p_{T}$ spectrum at Tevatron by
leading-order event generators}},  
Mod. Phys. Lett. A {\bf 25} (2010) 3047; 
arXiv:0907.5056.

\bibitem{Aad:2011gj}
ATLAS Collaboration, G.~Aad {\em et~al.}, {\it {Measurement of the
transverse momentum distribution of Z/$\gamma^{*}$ bosons in proton-proton
collisions at $\sqrt{s}$ = 7 TeV with the ATLAS detector}},  
Phys. Lett. B {\bf 705} (2011) 415; 
arXiv:1107.2381.

\bibitem{Sherstnev:2007nd}
A.~Sherstnev and R.~S. Thorne, {\it {Parton Distributions for LO Generators}},
Eur. Phys. J. C {\bf 55} (2008) 553; 
arXiv:0711.2473.

\bibitem{Lai:2009ne}
H.-L. Lai {\em et~al.}, {\it {Parton Distributions for Event Generators}},
JHEP {\bf 04} (2010) 035; 
arXiv:0910.4183.

\bibitem{Whalley:2005nh}
M.~R. Whalley, D.~Bourilkov, and R.~C. Group, {\it {The Les Houches Accord PDFs 
(LHAPDF) and Lhaglue}},  arXiv:hep-ph/0508110.

\bibitem{Pumplin:2002vw}
J.~Pumplin {\em et~al.}, {\it {New generation of parton distributions with
uncertainties from global QCD analysis}},  JHEP {\bf 07} (2002) 012; 
arXiv:hep-ph/0201195.

\bibitem{Martin:2009iq}
A.~Martin, W.~Stirling, R.~Thorne, and G.~Watt, {\it {Parton distributions for the LHC}},  
Eur. Phys. J. C {\bf 63} (2009) 189; 
arXiv:0901.0002.

\bibitem{Lai:2010vv}
H.-L. Lai {\em et~al.}, {\it {New parton distributions for collider physics}},
Phys. Rev. D {\bf 82} (2010) 074024; 
arXiv:1007.2241.

\bibitem{Abazov:2010kn}
D0 Collaboration, V.~M. Abazov {\em et~al.}, {\it {Measurement of the
normalized $Z / \gamma^{*} \rightarrow \mu^{+} \mu^{-}$ transverse momentum
distribution in $p\bar{p}$ collisions at $\sqrt{s} =$ 1.96 TeV}},  
Phys. Lett. B {\bf 693} (2010) 522; 
arXiv:1006.0618.

\bibitem{Abazov:2007nt}
D0 Collaboration, V.~M. Abazov {\em et~al.}, {\it {Measurement of the 
shape of the boson transverse momentum distribution in $p \bar{p} \rightarrow
Z / \gamma^{*} \rightarrow e^{+} e^{-}$ + $X$ events produced at $\sqrt{s}$ =
1.96 TeV}},  Phys. Rev. Lett. {\bf 100} (2008) 102002; 
arXiv:0712.0803.

\end{thebibliography}
\end{document}